\documentclass[12pt]{article}
\usepackage{mathrsfs}

\usepackage{fullpage}
\usepackage{amsfonts}
\usepackage{graphicx}
\usepackage{mathrsfs}
\usepackage{amsmath}
\usepackage{amssymb}
\usepackage{float}
\newtheorem{thm}{Theorem}
\newtheorem{lemma}{Lemma}

\def\var{\mathrm {var}}

\newcommand{\bm}{\boldsymbol}

\def\tr{\mathrm {tr}}

\def\D{{\bf D}}

\def\z{{\bf z}}

\def\I{{\bf I}}
\def\bmu{{\bm \mu}}

\def\X{{\bf X}}

\def\Y{{\bf Y}}
\def\BL{{\bf \Lambda}}

\def\tr{\mathrm {tr}}

\def\bth{{\bm\mu}}
\def\bms{{\bm\Sigma}}
\def\diag{\mathrm {diag}}

\def\cd{\overset{\mathcal{L}}{\longrightarrow}}

\title{\bf A Note on High Dimensional Two Sample Mean Test}
\author{Long Feng and Fasheng Sun\\
{\em \small Northeast Normal University}}
\date{}
\begin{document}
\maketitle

\begin{abstract}
In this paper, we propose a new scalar and shift transform invariant test statistic for the high-dimensional two-sample location test. The expectation of our test is exactly zero under the null hypothesis. And we allow the dimension could be arbitrary large. Theoretical results and simulation comparison show the good performance of our test.
\end{abstract}

\section{Introduction}\baselineskip 16.5pt
This article is concerned with the two-sample Behrens-Fisher problem
in high-dimensional settings. Assume that
$\{\X_{i1},\cdots,\X_{in_i}\}$ for $i=1,2$ are two independent
random samples with sizes $n_1$ and $n_2$, from $p$-variate
distributions $F({\bf x}-\bth_1)$ and $G({\bf x}-\bth_2)$ located at
$p$-variate centers $\bth_1$ and $\bth_2$. Denote $n=n_1+n_2$. We
wish to test
\begin{align}\label{ht}
H_0:\bth_1=\bth_2\ \ \mbox{versus}\ \  H_1:\bth_1\neq\bth_2,
\end{align}
where their covariances $\bms_1$ and $\bms_2$ are unknown. If $\bms_1=\bms_2$, the classic Hotelling's $T^2$ test is a nature
choice when the dimension is fixed and small. However, if the dimension is larger than the sample sizes, Hotelling's $T^2$ test can not work.
Recently, many efforts have been devoted to construct new test procedure under the high-dimensional settings. A nature method is replacing
the sample covariance matrix by the identity matrix (Bai and Saranadasa 1996, Chen and Qin 2010). However, those test statistics are not scalar-invariant.
Srivastava and Du (2008)  proposed a scalar-transformation-invariant test by replacing the sample covariance matrix with its diagonal matrix. And Srivastava, Katayama and Kano (2013) extend it to the unequal covariance case. However, the requirement of $p$ is a smaller order of $n^2$ is too restrictive to be used in high-dimensional settings. Feng, et al. (2014) propose another scalar-transformation-invariant test which allows the dimension being a smaller order of $n^3$.
Gregory, et al. (2014) proposed the generalized component test with $p=o(n^6)$. However, the requirement of $p$ being of the polynomial order of $n$ is too restrictive to be used in the ``large $p$ small $n$" situation. Park and Ayyala (2013) also propose a scalar-transformation-invariant test which allow the dimension could be arbitrary large. However, their test is not shift-invariant. Even each $\D^{-1}_{S_1^*(i,j)}$, $\D^{-1}_{S_2^*(i,j)}$, $\D^{-1}_{S_{12}^*(i,j)}$ are ratio-consistent estimator, the difference between these estimators are not ignorable. After some tedious calculation, we can show that
\begin{align*}
E(T_{PA})=\sum_{k=1}^p\frac{2n^{-2}\mu_k^2(\sigma_{1k}^2-\sigma_{2k}^2)^2}{(\kappa \sigma_{1k}^2+(1-\kappa)\sigma_{2k}^2)^3}(1+o(1))
\end{align*}
where the common vector $\bmu_1=\bmu_2=\bmu=(\mu_1,\cdots,\mu_p)$ and $n_1/n\to \kappa$. Thus, if the variances of the two samples are not all equal and the common vector is very large, $E(T_{PA})$ is not zero even under the null hypothesis. To overcome this issue, we propose a novel test statistic which is not only scalar-invariant but also shift-invariant. Under the null hypothesis, the expectation of our test statistic is exactly zero. There is no bias term in our test statistic. In addition, we also do not require the relationship between the dimension and the sample sizes. The dimension $p$ can be arbitrary large in this case. The asymptotic normality of the
proposed test can be derived under some very mild conditions similar to those in
Chen and Qin (2010).

The rest of the paper is organized as follow. In section 2, we propose the new test statistic and establish its the asymptotic normality. Simulation comparison is conducted in Section 3. We provide all the technical details in the appendix.

\section{Our test}
We now propose a new shift and scalar transformation invariant test statistic in the two sample test.
Define
\begin{align*}
T_n=\frac{1}{n_1(n_1-1)}\frac{1}{n_2(n_2-1)}\sum_{k=1}^p  \underset{i\not=j}{\sum^{n_1}\sum^{n_1}}
 \underset{s\not=t}{\sum^{n_2}\sum^{n_2}}\frac{(X_{1ik}-X_{2sk})(X_{1jk}-X_{2tk})}{\hat{\sigma}^2_{1k(i,j)}+\gamma\hat{\sigma}^2_{2k(s,t)}}
\end{align*}
where $\gamma=n_1/n_2$, $\hat{\sigma}^2_{1k(i,j)}$ is the sample variance of $\{X_{1lk}\}_{l=1}^{n_1}$ excluding $X_{1ik}$ and $X_{1jk}$. So does $\hat{\sigma}^2_{2k(s,t)}$. Because the numerator $(X_{1ik}-X_{2sk})(X_{1jk}-X_{2tk})$ is independent of the denominator $\hat{\sigma}^2_{1k(i,j)}+\gamma\hat{\sigma}^2_{2k(s,t)}$, thus,
\begin{align*}
E\left(\frac{(X_{1ik}-X_{2sk})(X_{1jk}-X_{2tk})}{\hat{\sigma}^2_{1k(i,j)}+\gamma\hat{\sigma}^2_{2k(s,t)}}\right)=&E((X_{1ik}-X_{2sk})(X_{1jk}-X_{2tk}))
E\left(\{\hat{\sigma}^2_{1k(i,j)}+\gamma\hat{\sigma}^2_{2k(s,t)}\}^{-1}\right)\\
=&(\mu_{1k}-\mu_{2k})^2E\left(\{\hat{\sigma}^2_{1k(i,j)}+\gamma\hat{\sigma}^2_{2k(s,t)}\}^{-1}\right)
\end{align*}
Unlike the three different estimators of $\sigma_{1k}^2+\gamma\sigma_{2k}^2$ for the three parts of the test statistic in Park and Ayyala (2013), we use the leave-two-out sample variance for each
numerator. Now, $E\left(\{\hat{\sigma}^2_{1k(i,j)}+\gamma\hat{\sigma}^2_{2k(s,t)}\}^{-1}\right)$ is exactly same for each numerator. Then,
\begin{align*}
E(T_n)=\sum_{k=1}^p (\mu_{1k}-\mu_{2k})^2E\left(\{\hat{\sigma}^2_{1k(i,j)}+\gamma\hat{\sigma}^2_{2k(s,t)}\}^{-1}\right).
\end{align*}
Thus, under the null hypothesis $H_0$, $E(T_n)$ is exactly zero. Furthermore, under the Conditions (C1)-(C3) stated next, we can show that
\begin{align*}
E(T_n)=&||\BL(\bmu_1-\bmu_2)||^2+o(\sqrt{\var(T_n)})\\
\var(T_n)=&\Bigg\{\frac{2}{n_1(n_1-1)}\tr((\BL \bms_1 \BL)^2)+\frac{2}{n_2(n_2-1)}\tr((\BL \bms_2 \BL)^2)\\
&+\frac{4}{n_1n_2}\tr(\BL \bms_1\BL^2\bms_2\BL)\Bigg\}(1+o(1)).
\end{align*}
where $\BL=\diag\left\{(\sigma_{11}^2+\gamma\sigma_{21}^2)^{-1/2},\cdots, (\sigma_{1p}^2+\gamma\sigma_{2p}^2)^{-1/2}\right\}$.

To establish the asymptotic normality of $T_n$, we need
the following conditions. Assume, like Bai and Saranadasa (1996)
and Chen and Qin (2010) did, $\X_{ij}$'s come from the following
multivariate model:
\begin{align}\label{cqm}
\X_{ij}={\bf \Gamma}_i \bm \z_{ij}+\bth_i \quad \text{for}
\quad j=1,\cdots,n_i,\ i=1,2,
\end{align}
where each ${\bf \Gamma}_i$ is a $p\times m$ matrix for some $m\ge
p$ such that ${\bf \Gamma}_i {\bf \Gamma}_i^{T}={\bf \bms}_i$, and
$\{\bm \z_{ij}\}_{j=1}^{n_i}$ are $m$-variate independent and
identically distributed random vectors such that
\begin{align}\label{chends}
\begin{array}{c}
 E(\z_i)=0, \ \var(\z_i)=\I_m,~E(z_{il}^{4})=3+\Delta,\Delta>0,~ E(z_{il}^{8})=m_8\in (0,\infty),\\
  E(z_{ik_1}^{\alpha_1}z_{ik_2}^{\alpha_2}\cdots
  z_{ikq}^{\alpha_q})=E(z_{ik_1}^{\alpha_1})E(z_{ik_2}^{\alpha_2})\cdots
  E(z_{ikq}^{\alpha_q}),
 \end{array}
\end{align}
for a positive integer $q$ such that $\sum_{k=1}^q\alpha_k\leq 8$ and $k_1\neq k_2\cdots\neq
k_q$.  The data structure  generates a rich collection
of $\X_i$ from $\z_i$ with a given covariance. Additionally, we need
the following conditions: as $n,p \to \infty$
\begin{itemize}
\item[(C1)] $n_1/(n_1+n_2)\to \kappa \in (0,1)$.
\item[(C2)] $\tr\left(\BL\bms_i\BL^2\bms_j\BL^2\bms_l\BL^2\bms_h\BL\right)=o(\tr^2\{(\BL \bms_1 \BL+\BL \bms_2 \BL)^2\})$ for $i,j,l,h=1$ or $2$.
\item[(C3)] $(\bm \mu_1-\bm \mu_2)^{T}\BL^2  \bms_i \BL^2(\bm \mu_1-\bm \mu_2)=o(n^{-1}\tr((\BL \bms_1 \BL+\BL \bms_2\BL)^2))$, for $i=1,2$. $((\bmu_1-\bmu_2)^T\BL(\bmu_1-\bmu_2))^2=o(n^{-1}\tr((\BL \bms_1 \BL+\BL \bms_2\BL)^2))$.
\end{itemize}

The following theorem establishes the asymptotic null distribution
of $T_n$.
\begin{thm}
Under Conditions (C1)--(C3), as $p,n \to \infty$,
\[\frac{T_n-E(T_n)}{\sqrt{\var(T_n)}}\cd N(0,1).\]
\end{thm}
Then, in order to formulate a testing procedure based on Theorem 1, we need to estimate the traces terms in $\var(T_n)$. Here, we adopt the following ratio-consistent estimators in Feng et al. (2014):
\begin{align*}
\widehat{\tr((\BL \bms_s \BL)^2)}=\frac{1}{2P_{n_s}^4}\sum^{*}
&(\X_{si_1}-\X_{si_2})^{T}\D_{s(i_1,i_2,i_3,i_4)}^{-1}(\X_{si_3}-\X_{si_4})\\
&\times(\X_{si_3}-\X_{si_2})^{T}\D_{s(i_1,i_2,i_3,i_4)}^{-1}(\X_{si_1}-\X_{si_4}),
\end{align*}
$s=1,2$, and
\begin{align*}
\widehat{\tr(\BL
\bms_1\BL^2\bms_2\BL)}=\frac{1}{4P_{n_1}^2P_{n_2}^2}\underset{i_1\not=i_2}{\sum^{n_1}\sum^{n_1}}\underset{i_3\not=i_4}{\sum^{n_2}\sum^{n_2}}
\left(
(\X_{1i_1}-\X_{1i_2})^{T}\D_{(i_1,i_2,i_3,i_4)}^{-1}(\X_{2i_3}-\X_{2i_4})\right)^2,
\end{align*}
where
\begin{align*}
\D_{1(i_1,i_2,i_3,i_4)}&=\diag(\hat{\sigma}_{11(i_1,i_2,i_3,i_4)}^2+\gamma\hat{\sigma}^2_{21},\cdots,\hat{\sigma}_{1p(i_1,i_2,i_3,i_4)}^2+\gamma\hat{\sigma}^2_{2p}),\\
\D_{2(i_1,i_2,i_3,i_4)}&=\diag(\hat{\sigma}_{11}^2+\gamma\hat{\sigma}^2_{21(i_1,i_2,i_3,i_4)},\cdots,\hat{\sigma}_{1p}^2+\gamma\hat{\sigma}^2_{2p(i_1,i_2,i_3,i_4)}),\\
\D_{(i_1,i_2,i_3,i_4)}&=\diag(\hat{\sigma}_{11(i_1,i_2)}^2+\gamma\hat{\sigma}^2_{21(i_3,i_4)},\cdots,\hat{\sigma}_{1p(i_1,i_2)}^2+\gamma\hat{\sigma}^2_{2p(i_3,i_4)}),
\end{align*}
and $\hat{\sigma}_{sk(i_1,\cdots,i_l)}^2$ is the $s$-th sample
variance after excluding $X_{si_j}$, $j=1,\cdots,l$, $s=1,2$,
$l=2,4$, $k=1,\cdots,p$. Through this article, we use
$\sum\limits^{*}$ to denote summations over distinct indexes. For
example, in $\widehat{\tr((\BL \bms_1 \BL)^2)}$, the summation is
over the set $\{i_1\not=i_2\not=i_3\not=i_4\}$, for all
$i_1,i_2,i_3,i_4\in\{1,\cdots,n_1\}$ and $P_n^m={n!}/{(n-m)!}$.

As a consequence, a ratio-consistent estimator of $\var(T_n)$ under $H_0$ is
\begin{align*}
\hat{\sigma}_n^2\doteq
\widehat{\var(T_n)}=\Bigg\{\frac{2}{n_1(n_1-1)}\widehat{\tr((\BL
\bms_1 \BL)^2)}&+\frac{2}{n_2(n_2-1)}\widehat{\tr((\BL \bms_2
\BL)^2)}\\
&+\frac{4}{n_1n_2}\widehat{\tr(\BL
\bms_1\BL^2\bms_2\BL)}\Bigg\}.
\end{align*}
This result suggests rejecting $H_0$ with $\alpha$ level of
significance if $T_n/\hat{\sigma}_n>z_{\alpha}$, where
$z_{\alpha}$ is the upper $\alpha$ quantile of $N(0,1)$.

\section{Simulation}
Here we report a simulation study designed to evaluate the performance of our proposed test (abbreviated as FS). We compare our tests with the method proposed by Chen and Qin (2010) (abbreviated as CQ), and Srivastava, Katayama and Kano (2013) (abbreviated as SKK), Park and Ayyala (2013) (abbreviated as PA) under the unequal covariance matrices assumption.  We
consider the following moving average model as Chen and Qin (2010):
\begin{align*}
X_{ijk}=\rho_{i1}Z_{ij}+\rho_{i2}Z_{i(j+1)}+\cdots+\rho_{iT_i} Z_{i(j+T_i-1)}+\mu_{ij}
\end{align*}
for $i=1,2$, $j=1,\cdots,n_i$ and $k=1,\cdots,p$ where $\{Z_{ijk}\}$
are, respectively, i.i.d. random variables. Consider two
scenarios for the innovation $\{Z_{ijk}\}$: (Scenario I) all the
$\{Z_{ijk}\}$ are from $N(0,1)$; (Scenario II) the first half
components of $\{Z_{ijk}\}_{k=1}^p$ are from centralized Gamma(4,1)
so that it has zero mean, and the rest half components are from
$N(0,1)$. The coefficients $\{\rho_{il}\}_{l=1}^{T_i}$ are
generated independently from $U(2,3)$ and are kept fixed once
generated through our simulations. The correlations among $X_{ijk}$
and $X_{ijl}$ are determined by $|k-l|$ and $T_i$. We choose $T_1=3$,
and $T_2=4$ to generate different covariances of $\X_i$.

We examine the empirical sizes and the estimation efficiency of tests. Under the null hypothesis, the components of common vector $\bmu_1=\bmu_2=\bmu_0=(\mu_1,\cdots,\mu_p)$ are generated from $U(0,\lambda)$. The sample sizes are $n_1=n_2=15$. First, we consider the impact of dimension. We fix $\lambda=10$ and consider six dimensions $p=25,50,100,200,400,800$. We summarize simulation results by using the mean-standard deviation-ratio (MDR) $E(T)/\sqrt{\var(T)}$ and the variance ratio (VR) $\widehat{\var(T)}/\var(T)$. Since the explicit form of $E(T)$ and $\var(T)$ is difficult to calculate, we estimate them by simulation. Figure 1 reports the MDR, VR and empirical sizes of these four tests with different dimensions. We observe that MDR and VR of SKK test are larger than zero and one when the dimension becomes larger. It is not strange because SKK must require the dimension is a smaller order of $n^2$. Second, we consider the impact of common shifts.  We fix the dimension $p=800$ and consider five common shifts $\lambda=10,20,30,40,50$. Figure 2 reports the MDR, VR and empirical sizes of these four tests with different common shifts. The MDR and VR of PA test become larger when the common shifts is larger. It further demonstrate that PA test is not shift-invariant. In contrast, the MDR and VR of our test is approximately zero and one, respectively. And then, we can control the empirical size very well. However, the empirical sizes of the other three tests deviate from the nominal level in most cases.

\begin{figure}[ht]
\begin{center}
\includegraphics[width=16.0cm,height=12cm]{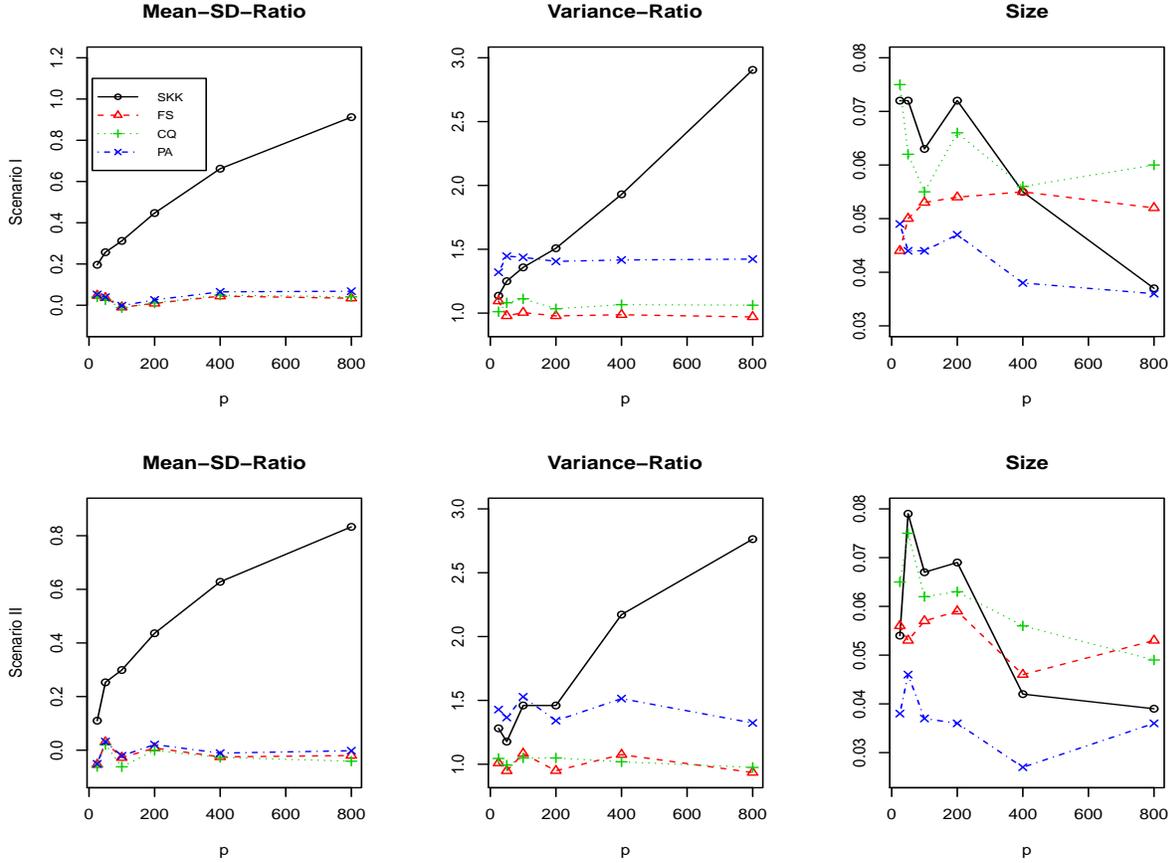}\vspace{-0.3cm}
\end{center}
\caption{\small  The MDR, VR and empirical sizes of tests with different dimensions.}\vspace{-0.5cm} \label{f2}
\end{figure}

\begin{figure}[ht]
\begin{center}
\includegraphics[width=16.0cm,height=12cm]{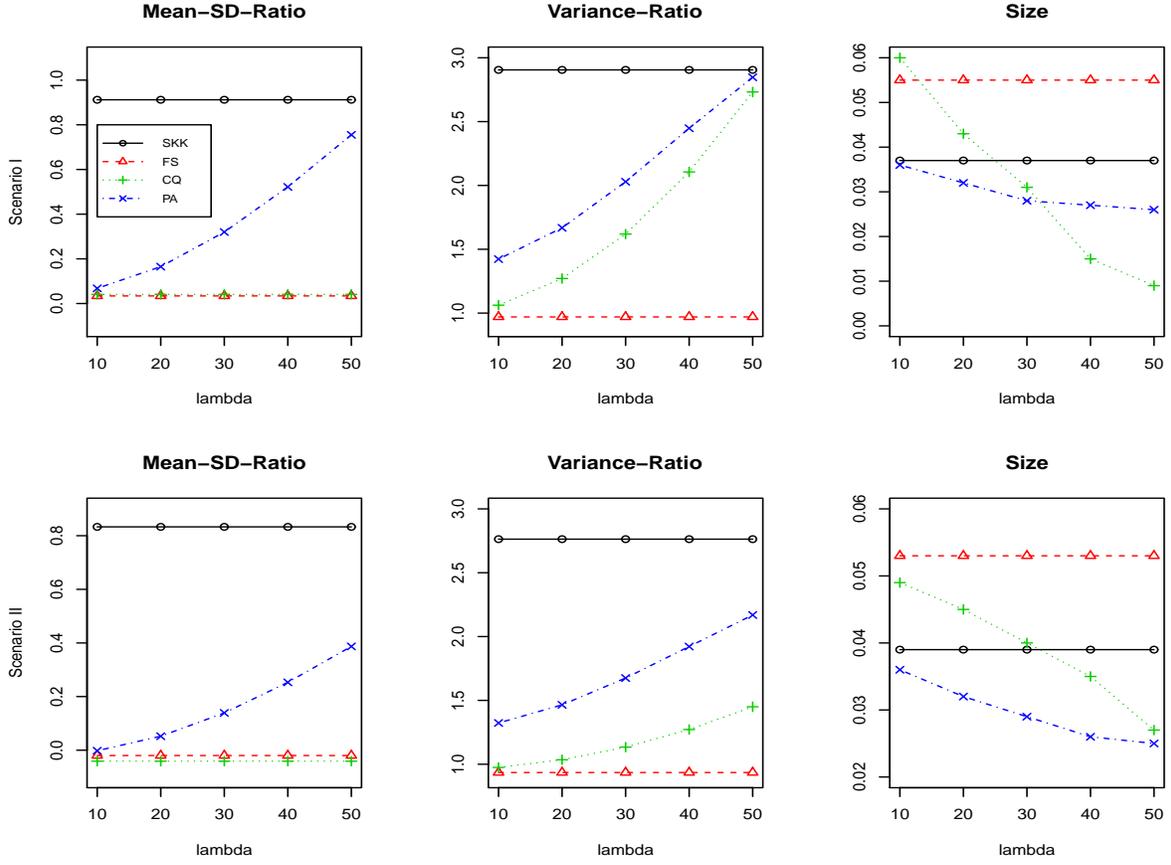}\vspace{-0.3cm}
\end{center}
\caption{\small  The MDR, VR and empirical sizes of tests with different $\lambda$.}\vspace{-0.5cm} \label{f2}
\end{figure}

Next, we compare the power of all these tests.
Here, we only report the case $n_1=n_2=15, p=800$. For the alternative hypothesis, $\bmu_1=\bmu_0+\bmu$ and $\bmu_2=\bmu_0$ where $\bmu_0$ are generated as above. We choose $\bm \mu$ in two scenarios: (Case A) one allocates all of the components of
equal magnitude to be nonzero; (Case B) the other allocates randomly half of
components of equal magnitude to be nonzero. To make the power
comparable among the configurations of $H_1$, we set $\eta:=||\bm
\mu_1-\bm \mu_2||^2/\sqrt{\tr(\bms_1^2)+\tr(\bms_2^2)}=0.15,0.2,0.25,0.3,0.35$ throughout the simulation. Figure 3 reports the empirical power of these four tests. Under Scenario II, CQ is less powerful than the other three tests because it is not scalar-invariant. Furthermore, our test performs better than SKK and PA tests in all cases. All these results together suggest that
the newly proposed FS test is scale and shift invariant and quite efficient and robust in
testing the equality of locations, and particularly useful when the variances of
components are not equal and the dimension is ultra-high.

\begin{figure}[ht]
\begin{center}
\includegraphics[width=16.0cm,height=12cm]{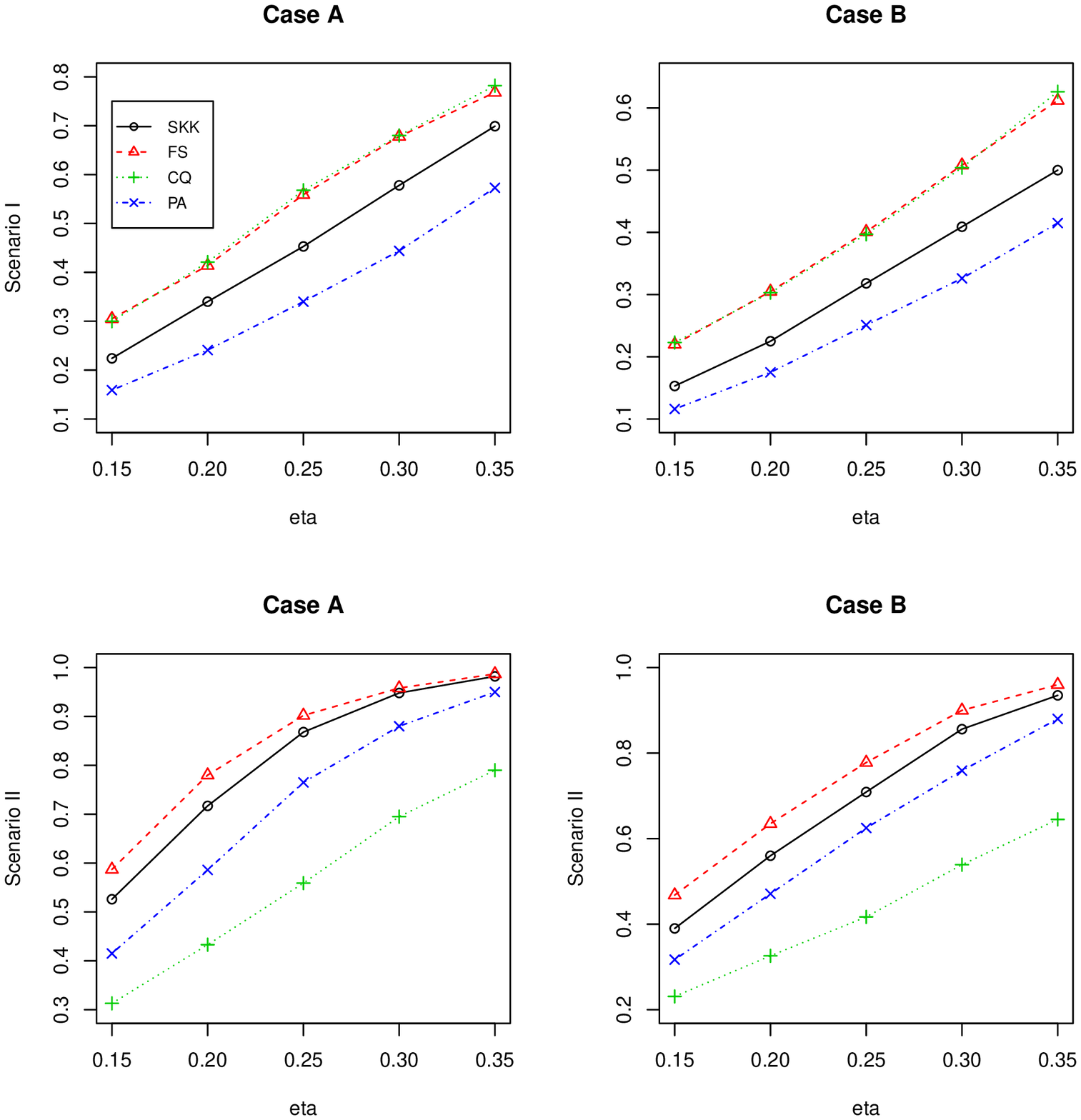}\vspace{-0.3cm}
\end{center}
\caption{\small  The power of tests with different $\eta$ when $n_1=n_2=15, p=800$.}\vspace{-0.5cm} \label{f2}
\end{figure}

\vspace{-0.3cm}
\appendix
\section*{Appendix: Proof of Theorem 1}

Firstly, after some tedious calculations, we decompose $T_n$ into
two parts, that is,
\begin{align*}
T_n=&\frac{1}{n_1(n_1-1)}\frac{1}{n_2(n_2-1)}\sum_{k=1}^p \underset{i\not=j}{\sum^{n_1}\sum^{n_1}}\underset{s\not=t}{\sum^{n_2}\sum^{n_2}}\frac{(X_{1ik}-X_{2sk})(X_{1jk}-X_{2tk})}{{\sigma}^2_{1k}+\gamma{\sigma}^2_{2k}}\\
&+\frac{1}{n_1(n_1-1)}\frac{1}{n_2(n_2-1)}\sum_{k=1}^p \underset{i\not=j}{\sum^{n_1}\sum^{n_1}}\underset{s\not=t}{\sum^{n_2}\sum^{n_2}}(X_{1ik}-X_{2sk})(X_{1jk}-X_{2tk})\\
& ~~~~~~~~~~~~~~~~~~~~~~~~~~\times\left(\frac{1}{\hat{\sigma}^2_{1k(i,j)}+\gamma\hat{\sigma}^2_{2k(s,t)}}-
\frac{1}{{\sigma}^2_{1k}+\gamma{\sigma}^2_{2k}}\right)\\
\doteq & T_{n1}+T_{n2}
\end{align*}

Then, it is straightforward to see that
\begin{align*}
E(T_{n1})&=\sum_{k=1}^p\frac{(\mu_{1k}-\mu_{2k})^2}{\sigma_{1k}^2+\gamma\sigma_{2k}^2}=||\BL(\bm \mu_1-\bm \mu_2)||^2,\\
\var(T_{n1})&=\frac{2}{n_1(n_1-1)}\tr((\BL \bms_1 \BL)^2)+\frac{2}{n_2(n_2-1)}\tr((\BL \bms_2 \BL)^2)+\frac{4}{n_1n_2}\tr(\BL \bms_1\BL^2\bms_2\BL)\\
&+\frac{4}{n_1}(\bm \mu_1-\bm \mu_2)^{T}\BL^2\bms_1\BL^2(\bm
\mu_1-\bm \mu_2)+\frac{4}{n_2}(\bm \mu_1-\bm
\mu_2)^{T}\BL^2\bms_2\BL^2(\bm \mu_1-\bm \mu_2).
\end{align*}
\begin{lemma}
Under the same conditions as Theorem 1, as $p$ and $n \to \infty$,
\begin{align*}
\frac{T_{n1}-E(T_{n1})}{\sqrt{\var(T_{n1})}} \cd N(0,1).
\end{align*}
\end{lemma}
This lemma is a direct corollary of Theorem 1 in Chen and Qin
(2010).

Next, we only need to show that $T_{n2}=o_p(\sqrt{\var(T_{n1})})$. Define $Y_{ijk}=X_{ijk}-\mu_{ik}$, $i=1,2,j=1,\cdots,n_i,k=1,\cdots,p$.
\begin{align*}
T_{n2}=&\frac{1}{n_1(n_1-1)}\frac{1}{n_2(n_2-1)}\sum_{k=1}^p \underset{i\not=j}{\sum^{n_1}\sum^{n_1}}\underset{s\not=t}{\sum^{n_2}\sum^{n_2}}(Y_{1ik}-Y_{2sk})(Y_{1jk}-Y_{2tk})\\
& ~~~~~~~~~~~~~~~~~~~~~~~~~~\times\left(\frac{1}{\hat{\sigma}^2_{1k(i,j)}+\gamma\hat{\sigma}^2_{2k(s,t)}}-
\frac{1}{{\sigma}^2_{1k}+\gamma{\sigma}^2_{2k}}\right)\\
&+\frac{2}{n_1n_2}\sum_{k=1}^p \sum_{i=1}^{n_1}\sum_{s=1}^{n_2}(Y_{1ik}-Y_{2sk})(\mu_{1k}-\mu_{2k})\\
& ~~~~~~~~~~~~~~~~~~~~~~~~~~\times\left(\frac{1}{\hat{\sigma}^2_{1k(i,j)}+\gamma\hat{\sigma}^2_{2k(s,t)}}-
\frac{1}{{\sigma}^2_{1k}+\gamma{\sigma}^2_{2k}}\right)\\
&+\sum_{k=1}^p (\mu_{1k}-\mu_{2k})^2\left(\frac{1}{\hat{\sigma}^2_{1k(i,j)}+\gamma\hat{\sigma}^2_{2k(s,t)}}-
\frac{1}{{\sigma}^2_{1k}+\gamma{\sigma}^2_{2k}}\right)\\
\doteq & R_1+R_2+R_3
\end{align*}
Define $\BL_{(i,j,s,t)}=\diag\{(\hat{\sigma}_{11(i,j)}^2+\gamma\hat{\sigma}_{21(s,t)}^2)^{-1/2},\cdots,(\hat{\sigma}_{1p(i,j)}^2+\gamma\hat{\sigma}_{2p(s,t)}^2)^{-1/2}\}$.
\begin{align*}
R_1
=&\frac{1}{n_2(n_2-1)}\sum_{t\not=s}^{n_2}\sum_{s=1}^{n_2}\left(\frac{1}{n_1(n_1-1)}\sum^{n_1}_{j\not=i}\sum^{n_1}_{i=1}\Y_{1i}^T (\hat{\BL}_{(i,j,s,t)}^2-\BL^2)\Y_{1j}\right)\\
&+\frac{1}{n_1(n_1-1)}\sum^{n_1}_{j\not=i}\sum^{n_1}_{i=1}\left(\frac{1}{n_2(n_2-1)}\sum_{t\not=s}^{n_2}\sum_{s=1}^{n_2}\Y_{2s}^T (\hat{\BL}_{(i,j,s,t)}^2-\BL^2)\Y_{2t}\right)\\
&-\frac{2}{(n_1-1)(n_2-1)}\sum_{j\not=i}^{n_1}\sum_{s\not=t}^{n_2}\left(\frac{1}{n_1n_2}\sum_{i=1}^{n_1}\sum_{s=1}^{n_2}\Y_{1i}^T (\hat{\BL}_{(i,j,s,t)}^2-\BL^2)\Y_{2t}\right)
\end{align*}
By the Theorem 1 in Park and Ayyala (2013), we have
\begin{align*}
&E\left(\frac{1}{n_1(n_1-1)}\sum^{n_1}_{j\not=i}\sum^{n_1}_{i=1}\Y_{1i}^T (\hat{\BL}_{(i,j,s,t)}^2-\BL^2)\Y_{1j}\right)^2=O(n^{-3}\tr((\BL\bms_1\BL)^2))=o(\var(T_{n1}))\\
&E\left(\frac{1}{n_2(n_2-1)}\sum_{t\not=s}^{n_2}\sum_{s=1}^{n_2}\Y_{2s}^T (\hat{\BL}_{(i,j,s,t)}^2-\BL^2)\Y_{2t}\right)^2=O(n^{-3}\tr((\BL\bms_2\BL)^2))=o(\var(T_{n1}))\\
&E\left(\frac{1}{n_1n_2}\sum_{i=1}^{n_1}\sum_{s=1}^{n_2}\Y_{1i}^T (\hat{\BL}_{(i,j,s,t)}^2-\BL^2)\Y_{2t}\right)^2=O(n^{-3}\tr(\BL\bms_1\BL^2\bms_2\BL))=o(\var(T_{n1}))
\end{align*}
Thus, $R_1=o_p(\sqrt{\var(T_{n1})})$. Similarly,
\begin{align*}
R_2=&\frac{2}{n_2(n_2-1)(n_1-1)}\sum_{t\not=s}^{n_2}\sum_{s=1}^{n_2}\sum_{j\not=i}^{n_1}
\left(\frac{1}{n_1}\sum_{i=1}^{n_1}\Y_{1i}^T(\hat{\BL}_{(i,j,s,t)}^2-\BL^2)(\bmu_1-\bmu_2)\right)\\
&-\frac{2}{n_1(n_1-1)(n_2-1)}\sum_{j\not=i}^{n_1}\sum_{i=1}^{n_1}\sum_{t\not=s}^{n_2}
\left(\frac{1}{n_2}\sum_{s=1}^{n_1}\Y_{2s}^T(\hat{\BL}_{(i,j,s,t)}^2-\BL^2)(\bmu_1-\bmu_2)\right)
\end{align*}
By the proof of Theorem 2 in Park and Ayyala (2013), we have
\begin{align*}
E\left(\frac{1}{n_1}\sum_{i=1}^{n_1}\Y_{1i}^T(\hat{\BL}_{(i,j,s,t)}^2-\BL^2)(\bmu_1-\bmu_2)\right)^2=O(n^{-2}(\bmu_1-\bmu_2)^T\BL^2\bms_1\BL^2(\bmu_1-\bmu_2))\\
E\left(\frac{1}{n_2}\sum_{s=1}^{n_1}\Y_{2s}^T(\hat{\BL}_{(i,j,s,t)}^2-\BL^2)(\bmu_1-\bmu_2)\right)^2=O(n^{-2}(\bmu_1-\bmu_2)^T\BL^2\bms_2\BL^2(\bmu_1-\bmu_2))
\end{align*}
By the Condition (C3), we also have $R_2=o_p(\sqrt{\var(T_{n1})})$. And $E(R_3^2)=O(n^{-1}((\bmu_1-\bmu_2)^T\BL(\bmu_1-\bmu_2))^2)=o(\var(T_{n1}))$. Thus, we proof that $T_{n2}=o_p(\sqrt{\var(T_{n1})})$.


\vspace{0.5cm} \noindent{\small\bf References} \footnotesize
\baselineskip 7pt
\begin{description}
\item Anderson, T. W. 2003. { An Introduction to Multivariate
Statistical Analysis,} Hoboken, NJ: Wiley.

\item Bai Z. and Saranadasa, H.  1996.
Effect of high dimension: by an example of a two sample problem,
{ Statistica Sinica}, { 6}, 311--329.

\item Biswas, M. and Ghosh. A. K. 2014. A nonparametric two-sample test applicable to high dimensional data, { Journal of Multivariate Analysis}, { 123}, 160--171.

\item Cai, T., Liu, W. D. and Xia, Y. 2013. Two-sample test of high dimensional means under dependency,
{ Journal of the Royal Statistical Society: Series B}, {76}, 349--372.

\item Chen, L. S., Paul, D., Prentice, R. L. and Wang, P. 2011. A regularized Hotelling's $T^2$ test for
pathway analysis in proteomic studies, {Journal of the American Statistical Association}, { 106}, 1345--1360.

\item Chen, S. X. and Qin, Y-L. 2010. A two-sample test for high-dimensional data with applications to
gene-set testing,  {The Annals of Statistics},  38, 808--835.

\item Dempster, A. P. 1958. A high dimensional two sample significance test, { The Annals of Mathematical Statistics}, { 29}, 995--1010.

\item Feng, L., Zou, C., Wang, Z. and Zhu, L. 2014. Two sample Behrens-Fisher problem for high-dimensional data. {Statistica Sinica}, To appear.

\item Gregory, K. B., Carroll, R. J., Baladandayuthapani, V. and Lahiri, S. N. 2014.  A two-sample test for equality of means in high
dimension, {Journal of the American Statistical Association}, To appear.

\item Park, J. and Ayyala, D. N. 2013. A test for the mean vector in large dimension and small samples,
{ Journal of Statistical Planning and Inference}, { 143}, 929-943.

\item Srivastava, M. S. and Du, M. 2008. A test for the mean vector with
fewer observations than the dimension, { Journal of Multivariate Analysis},
{ 99}, 386--402.

\item Srivastava, M. S., Katayama, S. and Kano, Y. 2013. A two sample test in high dimensional data,
{ Journal of Multivariate Analysis}, {114}, 349--358.

\item Zhang, J. T. and Xu, J. F. 2009. On the $k$-sample Behrens-Fisher problem
for high-dimensional data, {Science in China Series A: Mathematics}, { 52},
1285--1304.
\end{description}

\end{document}